\documentclass{pazhbeng}
\usepackage{graphicx}
\usepackage{floatrow}
\usepackage{longtable}

\usepackage[hyperfootnotes=false,draft]{hyperref}

\usepackage[dvipsnames]{xcolor}
\usepackage{amsmath}
\usepackage{amssymb}
\usepackage{subcaption}

\usepackage{ulem}

\usepackage{color}

\def\be{\begin{equation}}
\def\ee{\end{equation}}
\def\ba{\begin{eqnarray}}
\def\ea{\end{eqnarray}}

\def\msun{M_\odot}

\def\ltsima{$\; \buildrel < \over \sim \;$}
\def\simlt{\lower.5ex\hbox{\ltsima}}
\def\gtsima{$\; \buildrel > \over \sim \;$}
\def\simgt{\lower.5ex\hbox{\gtsima}}

\definecolor{webgreen}{rgb}{0,.5,0}
\definecolor{webbrown}{rgb}{.6,0,0}
\definecolor{falured}{rgb}{0.5, 0.09, 0.09}

\definecolor{darkblue}{rgb}{0.1, 0.1, 0.6}

\definecolor{darkgreen}{rgb}{0.1, 0.4, 0.0}

\begin{document}

\journalinfo{2026}{52}{1}{1}[12]
\UDK{524.77}

\title{Destruction of the interstellar dust \\by a supernova}

\author{E.O. Vasiliev
 \email{eugstar@mail.ru}
 \addresstext{1}{Lebedev Physical Institute, Russian Academy of Sciences, 53 Leninsky Avenue, Moscow 119991 Russia}
}
\submitted{}
\begin{abstract}
Destruction of the interstellar dust proceeds primary behind supernova shocks. The previous estimates of the mass of the interstellar dust destroyed in the SN remnant do not take into account the physical properties of the ambient medium. Here we consider how some parameters, i.e. gas density and metallicity, can influence the destruction of the interstellar dust. We show that there are two regimes of the interstellar dust grains destruction in SN remnants: rapid and almost complete in compact low-mass SN remnants expanding in dense medium, and gradual and weak destruction in massive remnants evolving in the low-dense environment. When time for thermal sputtering is close to the dynamical one, i.e. to the SN remnant age, the mass of the interstellar dust destroyed in the SN remnant reaches its maximum value. We find that change of the ambient gas density results in the reduction of the dust mass logarithmically. We argue that dust cooling suppresses the interstellar dust destruction up to a factor of 1.6 by mass. This factor decreases for higher density of the ambient medium. We found that the dust mass depends linearly on gas metallicity as ${\rm log}~M_d \sim {\rm [Z/H]}$ or, in other words, on the dust-to-gas ratio as $M_d \sim \zeta_d$. In turn, the destruction efficiency is higher in low-metallicity environments due to relatively longer adiabatic phase. We point out that the mass of the interstellar dust destroyed per one SN in a high density environment of the high star formation regions like in local ultraluminous infrared and high-redshift massive galaxies is about several times smaller than that in the Milky Way diffuse medium.
\\
\\
\end{abstract}


\section{Introduction} 

Destruction efficiency of the interstellar dust by supernovae type II (SNII) shocks \citep[][and references therein]{Barlow1978,Draine1979b,Draine1979a} increases the tension between the theoretical estimates of  the dust production (by planetary nebulae, red giant and supergiants, carbon star winds and SNII) and observed dust masses in the Milky Way interstellar medium \citep[see][]{Draine2009}. A disbalance between the dust destruction and its replenishment is challenging not only for our Galaxy, but also for dust-rich submillimetre galaxies at low and high redshifts  \citep[e.g.][and others]{Morgan2003,Dwek2007,Michalowski2010a,Michalowski2010b,Santini2010,Gall2011,Valiante2011,Rowlands2014}.

The dust budget in galaxies is far from being understood \citep[see for recent discussion in][]{Mattsson2021,Kirchschlager2022,Peroux2023,Schneider2024,Shchekinov2025gal}. Several mechanisms were proposed to reduce the dust destruction efficiency, including dust-to-gas decoupling \citep[][]{Hopkins2016,Mattsson2019,Mattsson2022}, thermal instability in the SN ejecta \citep{Shchekinov2025jcap} and SN shock propagation in a clumpy medium \citep{Dedikov2025na,Scheffler2026}.

The characteristic dust lifetime in the Milky Way ISM against sputtering ranges in $t_{sp}\sim 3\times 10^8 - 3\times 10^9$ yr \citep[][]{McKee1989,Jones1994,Jones1994b,Slavin2015,Bocchio2014,Ginolfi2018,Micelotta2018,Ferrara2021}. 
These estimates are based on the simple argument that the interstellar gas with the dust-to-gas ratio $\zeta_d \sim 0.01$ is `cleaned' from the dust by SN during the Sedov (adiabatic) phase, thus, the total mass of the dust destroyed by a SN is \citep{McKee1989}
\be
 M_d \sim \zeta_d {E_{SN} \over (200 {\rm km~s^{-1})^2}} \sim 12.6~\msun.
\label{eq:md}
\ee
In this simple estimate there is no assumption about any properties of the ambient gas, except the dust-to-gas ratio. Nevertheless, this value is in a good agreement with the numerical calculations for the SN evolution in a homogeneous medium \citep[e.g.][]{Slavin2015,Dedikov2025na}. 

In general, the total mass of the dust destroyed by a SN should be dependent on the interplay between sputtering, cooling and dynamical timescales in a SN remnant, in other words, SN bubble \citep[see discussion in][]{Shchekinov2025jcap}. They however rely on gas density, metallicity, dust abundance, clumpiness of the medium and other parameters. For example, dust destruction inside the bubble\footnote{Owing to their inertia, the dust grains penetrate and form a layer far behind the SN forward shock front. At adiabatic SN expansion this layer is thicker than the SN shell \citep{vs2024}.} is inhibited in more inhomogeneous medium \citep{Dedikov2025na} and in higher-Mach turbulent medium \citep{Scheffler2026} owing to an intensive interaction of the SN forward shock with dense clumps and slices. While one may expect more effective destruction in denser gas. Thus, the interplay between the timescales requires more detailed consideration.

In this paper we study how the physical properties of an ambient medium can influence on the destruction of the interstellar dust in a bubble formed by a single SN explosion. The paper is organized as follows. Section~1 presents the theoretical rationale for a possibility of such influence. Section~2 describes the details of the numerical model. Section~3 presents the evolution of the swept-up interstellar dust in a SN remnant. In Section~4 we discuss possible consequences and in Section~5 we summarize our results.


\section{1.~Theoretical estimates} \label{sec:estimates}

Dust is efficiently destroyed by thermal sputtering in a gas with temperature $T\simgt 10^6$~K \citep{Draine1979b} on a timescale of 
\be
 t_{sp} = 10^5~ {\rm yr}~\left[ 1 + T_6^{-3}\right] {a_{0.1} \over n},
\label{eq:tsp}
\ee
where $T_6 = T/10^6$~K, $a_{0.1}$ is the grain size in units of 0.1~$\mu$m, $n$ is the number density of gas in units of 1~cm$^{-3}$. Gas at such temperature dominates by mass in a SN bubble, while which expands adiabatically. In this phase the radius of the bubble formed by a SN explosion with energy $E$ in the ambient gas with density $\rho = m_p n$ evolves as $r_s \sim (Et^2/\rho)^{1/5}$ and the gas density and temperature behind the forward shock can be found according to the Rankine-Hugoniot conditions with the adiabatic index $\gamma = 5/3$
\be
 n_s = 4 n, \hspace{1cm} T_s = {3\over 100} {m_p \over k_B} v_s^2,
\label{eq:ts}
\ee
where $v_s \sim (2/5) (E/\rho)^{1/5} t^{-3/5} \sim 2r_s/5t$. Using these values we estimate the sputtering time scale in the shell of the SN bubble at adiabatic phase. 

\begin{figure}
\includegraphics[width=8.5cm]{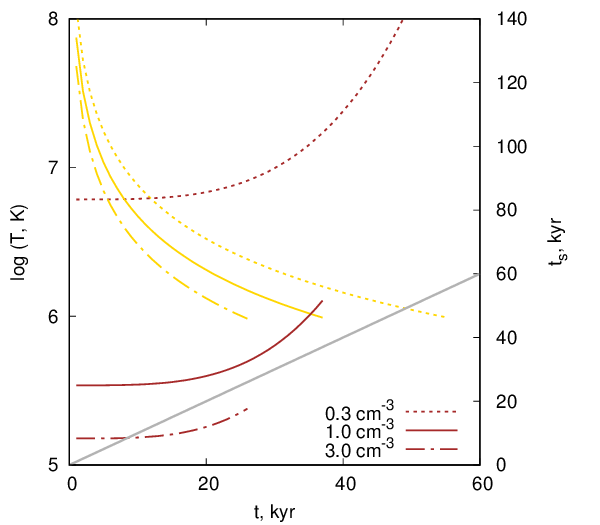}
\caption{
Gas temperature (yellow lines, left axis) and grain sputtering time (brown lines, right axis) in the SN bubble evolving adiabatically (the temperature is constrained by $10^6$~K) in the ambient medium with density $n = 3,\ 1$, and 0.3~cm$^{-3}$ (from bottom to top for yellow and brown lines, respectively). The grain size is $a_0 = 0.1\mu$m. The grey solid line depicts the dynamical time (right axis), i.e. the SN age.
}
\label{fig-t-est-evol}
\end{figure}

Fig.~\ref{fig-t-est-evol} illustrates the gas temperature and grain sputtering time in the SN bubble evolving  adiabatically. During this period the gas temperature follows eq.~\ref{eq:ts} and the sputtering remains efficient, i.e. $T_s\sim 10^6$~K. For the ambient gas density of $n_b=1$~cm$^{-3}$ the sputtering time (the middle brown line) is close to the SN age (grey solid line), i.e. the dynamical time, around $15-30$~kyr. During this period the swept-up interstellar grains are efficiently destroyed in the SN shell. An increase of gas density leads to shortening of the sputtering time. For $n_b=3$~cm$^{-3}$ it becomes shorter than the dynamical time at the SN age is $\sim 5$~kyr and remains below it until the onset of the radiative phase at $\sim 25$~kyr. This results in almost complete destruction of the grains in the shell. Decreasing the ambient gas density favours the surviving of the grains in the shell. For $n_b=0.3$~cm$^{-3}$ the sputtering time exceeds the dynamical time a factor of $3-4$ at the SN age $\sim 20-30$~kyr. Thus, grains are destroyed weakly.

The mass of the ambient gas swept-up by the SN forward shock scales as $M \sim n^{2/5}(t^{2/5})^3$. When the SN age becomes close to the cooling time, i.e. close to the end of adiabatic phase, the mass depends as $M\sim n^{-4/5}$. Hence, expanding in denser medium the SN shock sweeps smaller mass of the interstellar dust, but the major part of this dust is sputtered. In rarefied medium the SN shell becomes more massive at the end of adiabatic phase. In a SN shell the swept-up dust is destroyed too weakly. Therefore, there are two regimes of the interstellar dust destruction in SN remnants: {\it rapid and almost complete destruction of grains by compact less-massive SN remnants expanding in the dense medium and gradual and weak destruction in large massive remnants evolving in low-dense environment.}

In any case the mass of the interstellar dust swept-up by the SN shock depends on the interplay between sputtering, cooling and dynamical times. This picture becomes more complicated due to the inertia of the dust grains: they can penetrate far behind the shock front, while their velocities are close to the values of the  surrounding gas. Therefore, one more characteristic timescale, i.e. the stopping time, comes into play. Thus, the mass of the interstellar dust destroyed by the SN shock is expected to be dependent not only on SN energy, but also on the density and metallicity of the ambient gas, as well as, on grain size. In what follows, we consider these issues self-consistently using numerical approach for gas and dust dynamics.

\section{2.~The 3D model} \label{sec:model}

We carry out 3D Cartesian hydrodynamic simulations of the ISM under the action of a single SN explosion. We inject the mass and energy of a SN in a region of radius $r_0$, assuming commonly used values $M_{inj}=30~\msun$ and $E=10^{51}$~erg. The energy is injected in thermal form, although the results are not sensitive to the form of the injected energy \citep{Sharma2014}.

The gas distribution in the ambient (background) medium is set to be homogeneous with density $n_b = 0.1,\ 0.3,\ 1,\ 3$ and 10~cm$^{-3}$. The gas temperature in all models is $10^4$K.

Simulations are run with tabulated non-equilibrium cooling rates for a gas that cools isochorically from $10^8$ down to 10~K \citep{v11,v13}. The heating rate is chosen such as to stabilize the radiative cooling of the ambient gas.

The metallicity of the background gas is equal to the solar value ${\rm [Z/H]} = 0$ as a fiducial value. In the appropriate section we consider the set of values ${\rm [Z/H]} = -2,\ -1.5,\ -1,\ -0.5$ and $+0.5$. 

We consider single-sized interstellar dust $a_0 = 0.1\mu$m in our models below if otherwise is not specified. During the evolution, dust grains are destroyed and their sizes decrease depending on physical conditions in the ambient gas; the minimum size is set to $0.01\mu$m. We assume a dust-to-gas (DtG) mass ratio equal to $\zeta_b=10^{-2+{\rm [Z/H]}}$. 

We use our gas-dynamic code \citep{vns2015,vsn2017} based on the unsplit total variation diminishing (TVD) approach that provides high-resolution capturing of shocks and prevents unphysical oscillations, and the Monotonic Upstream-Centered Scheme for Conservation Laws (MUSCL)-Hancock scheme with the Haarten-Lax-van Leer-Contact (HLLC) method \citep[see e.g.][]{Toro2009} as approximate Riemann solver. This code has successfully passed the whole set of tests proposed in \cite{Klingenberg2007}. 

In order to follow the dynamics of the dust particles we have implemented the method \citep[see description and tests in Appendix A of][]{vs2024} similar to that proposed in \citet{Youdin2007}, \citet{Mignone2019} and \citet{Moseley2023}. The backward reaction of dust on to gas due to momentum transfer, work done by the drag force and the frictional heating from dust particles are also accounted for in order to ensure both dynamical and thermal self-consistency. We take into account the destruction of graphite, silicate and iron dust particles by both thermal (in a hot gas) and kinetic (due to a relative motion between gas and grains) sputtering \citep{Draine1979b}. We do not include shattering of grains in their each-other collisions, because the characteristic time of this process is probably to be longer the final time of the runs \citep[see the discussion in][]{Dedikov2025na}.

The cell sizes equal to 0.75, 0.5, 0.375, 0.25, 0.125~pc are adopted for the fiducial runs 'x1' with ambient gas number density $n_b = 0.1,\ 0.3,\ 1,\ 3$ and 10~cm$^{-3}$, respectively. We consider two another sets of runs with half '/2' and doubled 'x2' cell sizes for checking numerical convergence of our results. In all sets the resolution is sufficient to resolve the thermal (cooling) length estimated as $\lambda_t \sim 5~n^{-1} T_6$~pc.

The mass of dust in a medium is distributed between many dust `superparticles'. Each `superparticle' represents an ensemble of numerous physical grains of a single size. To follow the dust transport in a medium we set at least one `superparticle'  per computational cell. Each `superparticle' contains the total mass of the ensemble of single-sized dust particles in it. So the total dust mass in a cell is a sum of masses of `superparticles' inside a cell \citep[see for more details][]{vs2024}.

\section{3.~Dust destruction} \label{sec:dust} 

As mentioned above the problem depends on the interplay between sputtering, cooling and dynamical timescales in a SN remnant. Thus, we follow the SN bubble evolution until the age, when the dust destruction becomes insufficient, that means the SN age should be noticeably shorter than the sputtering timescale and longer than the cooling timescale. This allows us to be sure that the destruction of the interstellar dust has been ceased inside the bubble at late radiative phase.

\subsection{Gas and dust dynamics}

\begin{figure}
\includegraphics[width=8.5cm]{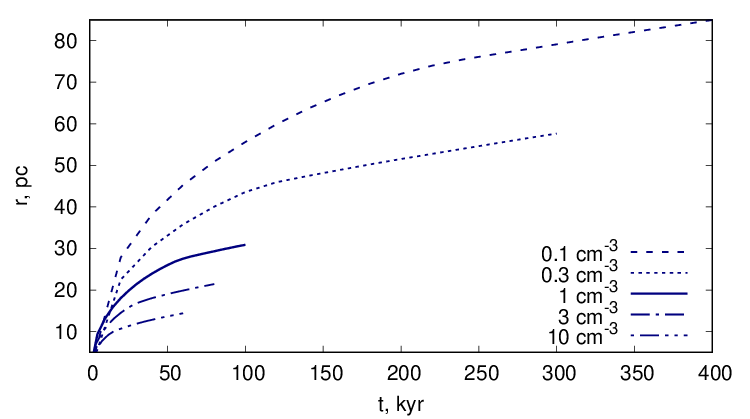}
\caption{
The radius of the SN bubble expanding in a gas with density $n_b = 0.1,\ 0.3,\ 1,\ 3$ and 10~cm$^{-3}$ from upper to lower lines, respectively. The solid line depicts the radius for 1~cm$^{-3}$.
}
\label{fig-rsn-evol}
\end{figure}

Fig.~\ref{fig-rsn-evol} shows the radius of the SN bubble expanding in a homogeneous medium with density $n_b$, solar metallicity and DtG ratio $\zeta_d = 0.01$. At early times the radius evolves adiabatically as $t^{2/5}$ \citep{Sedov1959}, after the energy losses become significant it scales approximately as $\sim t^{1/4}$ \citep{Oort1951,Blinnikov1982}. 

\begin{figure}
\includegraphics[width=9cm]{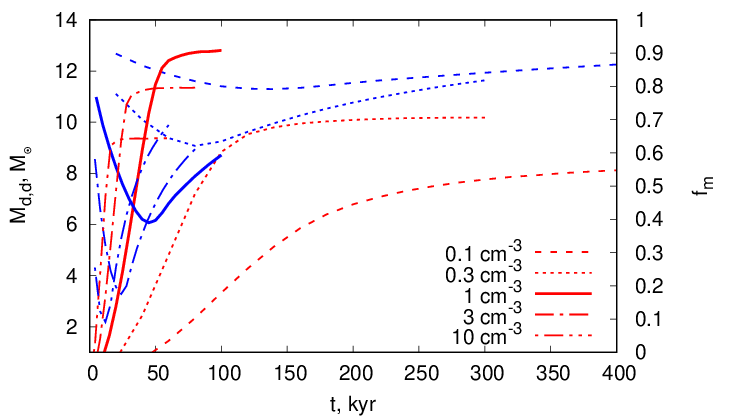}
\caption{
The mass of the interstellar dust (red lines, left axis) destroyed in the SN bubble expanding in a gas with density $n_b = 0.1,\ 0.3,\ 1,\ 3$ and 10~cm$^{-3}$ (from right to left red lines, respectively). The ’dust mass
survival’ fraction (blue lines, right axis) in the SN bubble. The solid lines depict the dependences for 1~cm$^{-3}$.
}
\label{fig-md-evol}
\end{figure}

During the expansion the interstellar dust is swept up with the ambient gas by the SN forward shock. Owing to the inertia, the dust grains penetrate and form a thick layer far behind the shock front \citep{vs2024}. In this layer the gas is hot and dense, therefore dust grains are subject to thermal sputtering. Fig.~\ref{fig-md-evol} shows the mass of the interstellar dust (left axis) destroyed in the SN bubble, i.e. the difference between the total mass of dust grains inside the SN bubble $M_d$ and this dust mass without taking into account grain destruction $M_{d,0}$, for various values of the ambient gas density. At early times, the mass of the dust destroyed in the SN bubble, $M_{d,d} = M_{d,0} - M_{d}$, grows linearly and saturating afterwards, when the gas temperature behind SN shock drops below $10^6$~K and the sputtering exhausts. One can see that the saturated value of the dust mass depends on the density non-monotonically: it varies from $\sim 9~\msun$ for $n_b=10$~cm$^{-3}$ to $\sim 8~\msun$ for $\sim 0.1$~cm$^{-3}$ with the maximum at $\sim 13~\msun$ for $n=1$~cm$^{-3}$. The maximum value is close to that obtained using the simple estimate by \citet{McKee1989}. 

Note, recently \citet{Scheffler2026} have concluded about increasing destruction efficiency of the interstellar dust under the action of a SN remnant expanding in denser medium. Using the postprocessing of the 3D numerical calculations for gas dynamics they included all processes impacting the dust evolution in the SN shocked ISM and obtained that the dust is destroyed less efficiently in a low-dense medium than in a high-dense one: for the average density $1$~cm$^{-3}$ about 27--57\% of the dust mass has been destroyed, while in the medium with $\sim 100$~cm$^{-3}$ this fraction grows to almost 73--87\%. One should pay attention that the analysis presented in \citep{Scheffler2026} has been constrained by short period of 10~kyr after the SN explosion and this period is the same for various average desnsities of the interstellar gas. At this age, in a low-dense gas an SN remnant expands with high enough velocity, which is sufficient for destroying dust that flow through the shock front. This can be suggested based on the final shock velocity at 10~kyr given by \citet{Scheffler2026}: $\sim 250-400$~km/s for the average density of $1$~cm$^{-3}$. From intuitive point of view it seems likely that the full MHD consideration at longer times would result to increasing the efficiency of dust destruction by SN in a low-density interstellar medium and, in general, to the results obtained in our work.

As one can see in Fig.~\ref{fig-md-evol}  the mass of the interstellar dust destroyed in the bubble grows at early period. For the same age, e.g. 10~kyr, the mass of the destroyed dust obviously grows with increase of ambient density as it has been obtained by \citep{Scheffler2026}. Afterwards though, when the SN shell becomes strongly radiative, it saturates at a certain level depending on the gas density (and metallicity as well). In a medium with density of $1$~cm$^{-3}$, approximately $\sim 1.5~\msun$ of dust has been destroyed at the age of 10~kyr, that is close to the value obtained in \citep{Scheffler2026}. However, at the onset of the radiative phase the mass of the destroyed dust reaches $\sim 9~\msun$ and continues to grow afterwards, while it saturates at $\sim 13~\msun$ (see red solid line Fig.~\ref{fig-md-evol}). The saturation is reached earlier for higher density of the ambient gas. For $n\sim 10$~cm$^{-3}$ that takes place at SN age of $\sim 15$~kyr and the dust mass attains $\sim 9~\msun$, while at $\sim 10$~kyr the mass is only about $6~\msun$ (left red line in Fig.~\ref{fig-md-evol}). This is the result of the ``interplay'' between the characteristic times of the problem, particularly between the sputtering and cooling times.

The overall picture becomes more complex in a clumpy medium with dispersion $\sigma\simgt 1$ for a log-normal density field, (for which $2\sigma$ fluctuations have the overdensity factor $\chi\simgt 10$) due to an additional time scale connected with the destruction of dense clumps. Dense clumpy gas shields dust particles against destruction and enhances radiative cooling of the hot gas in the SN bubbles, and thus increases the survived dust fraction \citep{Dedikov2025na}. A supply of the dust due to clump stripping leads to the onset of the saturation  later.

One can note that in our model we neglect shattering of grains in their each-other collisions and effects from magnetic fields. However, the characteristic time of this process is probably to be longer the final time of the runs \citep[see the discussion in][]{Dedikov2025na}. Anyway the influence of magnetic field on shattering efficiency is not apparently straightforward \citep{Moseley2023}.

\begin{figure}
\includegraphics[width=8.5cm]{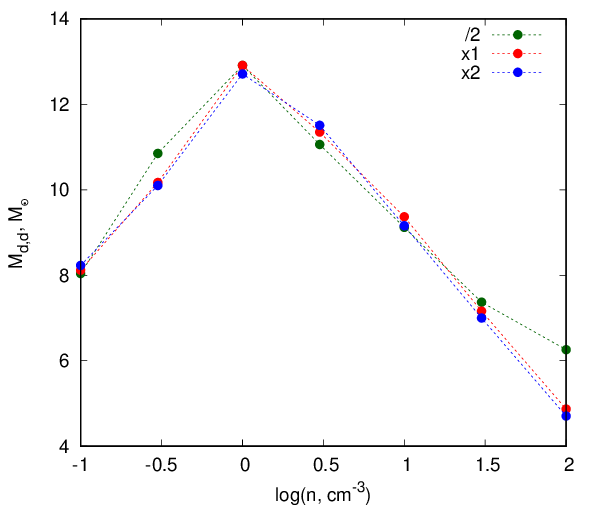}
\caption{
The mass of the interstellar dust destroyed in the bubble expanding in the ambient medium with density $n$.
The red line shows the runs with the fiducial spatial resolution ('x1'). The 
sets of runs with half ('/2') and doubled ('x2') cell sizes are depicted by green and blue lines, respectively.
}
\label{fig-md-sat}
\end{figure}

Fig.~\ref{fig-md-evol} presents the ’dust mass survival’ fraction (right axis) in the SN bubble, i.e. the ratio of the total mass of dust grains inside the SN bubble normalized to the dust mass without taking into account grain destruction, $f_m = M_d/M_{d,0}$. For the bubble evolved in denser medium the survival fraction is lower. This means that the major part of the swept-up interstellar dust is destroyed. For $n_b = 10$~cm$^{-3}$ almost 90\% of the mass of the dust in the bubble is sputtered till the end of adiabatic phase (see the minimum of the left blue line). Decreasing the ambient density rises the survival fraction, i.e. for $n_b = 0.1$~cm$^{-3}$ almost 80\% of the mass of the dust survives in the bubble (see the minimum of the right blue line). Note that the growth of the survival fraction after reaching the minimum is connected to abruptly ceased dust destruction after the radiative phase started and continous accumulation of the interstellar dust in the shell during further expansion of the SN remnant.

Fig.~\ref{fig-md-sat} depicts the mass of the interstellar dust destroyed in the bubble expanding in the ambient medium with density $n$. This value corresponds to the mass of the dust presented in Fig~\ref{fig-md-evol} after it reaches the saturation. Our calculations with the fiducial resolution ('x1') demonstrate reasonable numerical convergence with the results for the runs with half ('/2') and doubled ('x2') cell sizes. Thus, we can use the runs with the fiducial cell size in our further analysis. 

One can see that the mass of the interstellar dust destroyed in the bubble evolving in a rarefied medium with $n_b\simlt 1$~cm$^{-3}$ grows with increasing gas density approximately as $4.5~{\rm log}~n_b$. At $n_b\sim 1$~cm$^{-3}$ this mass reaches the maximum at $\sim 13~\msun$ and after that it decreases as $-4.1~{\rm log}~n_b$. Here we extend our set of models to higher ambient gas density. So, only $\sim 5~\msun$ of the interstellar dust is destroyed by a SN remnant evolving in a gas with $n_b \sim 100$~cm$^{-3}$. This value is more than a factor of two smaller than the estimate obtained by using the simple approach \citep{McKee1989}.

\begin{figure}
\includegraphics[width=8.5cm]{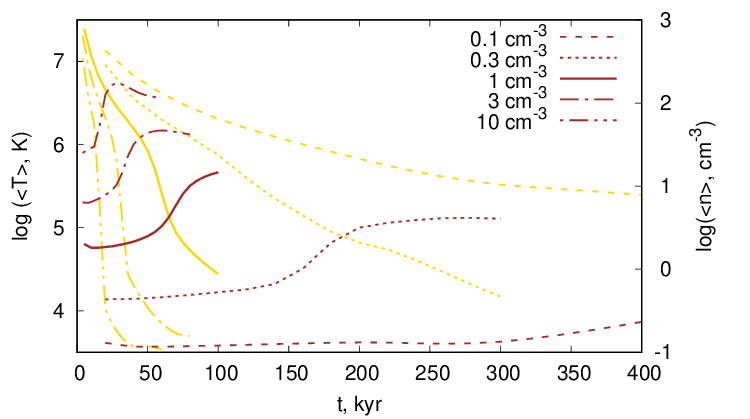}
\caption{
Gas temperature (yellow lines, left axis) and density (brown lines, right axis) averaged over the dust mass inside the SN bubble evolving in the ambient medium with density $n_b = 0.1,\ 0.3,\ 1,\ 3$ and 10~cm$^{-3}$ (from right to left for yellow lines and from bottom to top for brown lines, respectively). 
}
\label{fig-t-d-evol}
\end{figure}

\begin{figure}
\includegraphics[width=8.5cm]{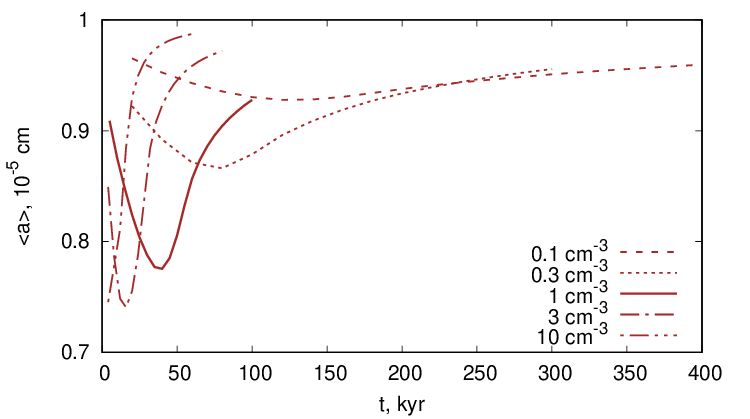}
\caption{
The mass-averaged grain sizes inside the SN bubble evolving in the ambient medium with density $n_b = 0.1,\ 0.3,\ 1,\ 3$ and 10~cm$^{-3}$ (lines from top-right to bottom-left, respectively).
}
\label{fig-a-evol}
\end{figure}

\begin{figure}
\includegraphics[width=8.5cm]{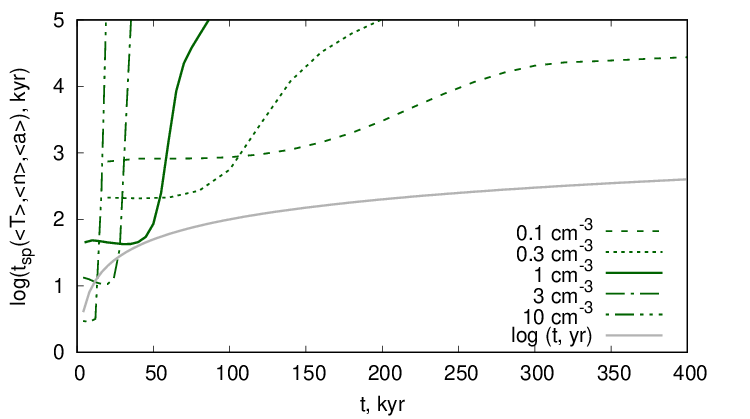}
\caption{
The sputtering time for the mass-averaged temperature and density of the gas (Fig.~\ref{fig-t-d-evol}) and grain size (Fig.~\ref{fig-a-evol}) for the SN bubble evolving in the ambient medium with density $n_b = 0.1,\ 0.3,\ 1,\ 3$ and 10~cm$^{-3}$ (lines from top-right to bottom-left, respectively). 
The grey solid line depicts the dynamical time, i.e. the SN age.
}
\label{fig-tsput-evol}
\end{figure}

The interstellar dust particles cross the shock front with zero velocity. Due to the interaction with gas they are accelerated up to the velocity of the surrounding gas flow within the stopping time scale \citep{Epstein1924,Baines1965}. In this process they can penetrate far behind the shock front \citep{vs2024}. They do not fill the entire hot SN bubble and are located in a thick layer behind the SN forward shock. In general, this layer contains the swept-up interstellar material and is separated from the SN ejecta by the contact discontinuity \citep[e.g.][]{Truelove1999}. The radial profiles of gas and dust densities in this layer are far from the flat distributions \citep[see e.g. Fig.~1 in][]{vs2024}. Therefore, we take into account those parts of the SN bubble, where there is the swept-up interstellar dust (more exactly, we only consider the numerical cells, where the dust superparticles are located). We calculate the gas temperature averaged with the mass of such dust: $\langle T\rangle = \int T_{gas} m_d dm / \int m_d dm$, where $m_d$ is the mass of dust superparticle, and the integration is over all superparticles located inside the SN bubble. The same is applied to the gas density and grain size. The (dust-)mass-averaged gas temperature, gas density and grain size are shown in Figs.~\ref{fig-t-d-evol} and \ref{fig-a-evol}. For these values we find the sputtering time $\langle t_{sp}\rangle(\langle T\rangle, \langle n\rangle, \langle a\rangle)$, these values are presented in Fig.~\ref{fig-tsput-evol}. For comparison, the grey solid line depicts the dynamical time, i.e. the SN age.

One can note that the sputtering time $\langle t_{sp}\rangle$ for the SN remnant evolving in the medium dense as $n_b \simgt 3$~cm$^{-3}$  becomes shorter than dynamical time even at early times (Fig.~\ref{fig-tsput-evol}).
While the gas temperature remains higher $10^6$~K (Fig.~\ref{fig-t-d-evol}), the swept-up dust is efficiently destroyed. The ratio $M_d/M_{d,0}$ reaches $\sim 0.2-0.3$ (see the right axis in Fig.~\ref{fig-md-evol}), that is a result of decreasing the mass-averaged grain size $\langle a \rangle$ about a factor of 0.75 during several thousands of years of the evolution (Fig.~\ref{fig-a-evol}). After the gas temperature drops below $10^6$~K, the sputtering time raises abruptly and the dust destruction has been ceased although the gas density is increased about an order of magnitude (Fig.~\ref{fig-t-d-evol}). 

A decrease of gas density in the ambient medium to $n_b\sim 1$~cm$^{-3}$ results in the sputtering time becoming close to the dynamical one (Fig.~\ref{fig-tsput-evol}), while the SN remnant expands adiabatically. In such gas the grain destruction rate drops proportionally to $n_b$, but the period, when the gas temperature remains as high as $10^6$~K, gets longer (see Fig.~\ref{fig-t-d-evol}). Thus, the mass-averaged grain size drops almost the same factor $\sim 0.78$ (see solid line in Fig.~\ref{fig-a-evol}), but this is reached on a longer timescale. This maximizes the mass of the interstellar dust destroyed in the SN bubble (Fig.~\ref{fig-md-sat}).

In a more rarefied ambient medium the sputtering time exceeds the dynamical one by more than a factor of several  during the whole evolution (Fig.~\ref{fig-tsput-evol}). The grains inside the bubble are destroyed less efficiently, their size decreases only  by a few percents (see the upper line in Fig.~\ref{fig-a-evol}). 

Based on Figs.~\ref{fig-a-evol} and Fig.~\ref{fig-tsput-evol} one can conclude about {\it a difference between dust destruction efficiency in a SN remnant evolving in high and low dense ambient medium}. At densities as high as $n> 1$~cm$^{-3}$ there is a rapid destruction of dust, while at $n< 1$~cm$^{-3}$ the grains are sputtered gently. When dust destruction is turned off, SN remnant evolving in high density environment is compact and low-massive, while it becomes large both in size and mass in case of low-density ambient medium. Therefore, these results confirm our theoretical arguments found in Sec.~1.

\subsection{Dependence on grain size}

\begin{figure}
\includegraphics[width=8.5cm]{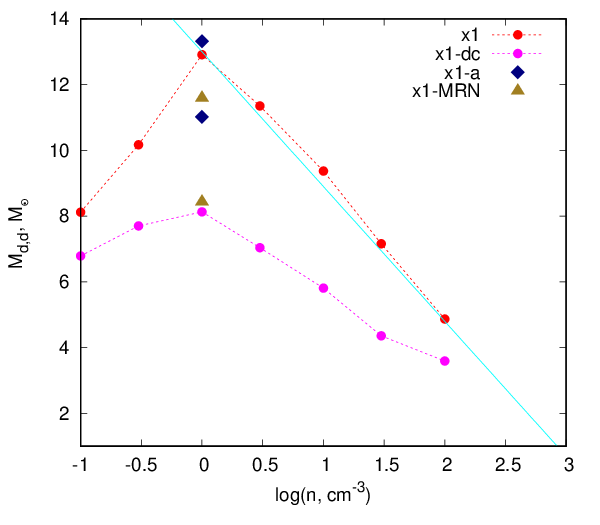}
\caption{
The same as in Fig.~\ref{fig-md-sat}. The red line with circles shows the runs with the fiducial spatial resolution as in Fig.~\ref{fig-md-sat}. The magenta line with circles depicts the same models, but taking into account dust cooling. The upper and lower diamonds represent the mass of the dust with initial size 0.075 and 0.125~$\mu$m, respectively, in the SN evolving in the medium with density $n=1$~cm$^{-3}$ without dust cooling. The triangle symbols show the mass of the interstellar dust with initial power-law distribution in range $0.003-0.3~\mu$m with index $-3.5$ destroyed in the SN remnant evolving in the medium with $n=1$~cm$^{-3}$: upper triangle is for the run without dust cooling, lower triangle depicts the run with it.
}
\label{fig-md-sat-v}
\end{figure}

In our runs we assume single-sized interstellar dust with initial value equal to $a_0 = 0.1\mu$m. The sputtering timescale is proportional to the grain size, and this changes the interrelation between the characteristic times of the problem and can have important consequences on the results. Fig.~\ref{fig-md-sat-v} shows the dust mass values  for grains with initial size 0.075 and 0.125~$\mu$m (upper and lower diamonds, respectively) in the SN evolving in the medium with $n=1$~cm$^{-3}$. It is apparent that the mass is decreased for larger grain size. 

Certainly, this is a simplification of the grain size distribution. There are more observationally motivated distribution functions \citep{Mathis1977,Jones1996,Weingartner2001,Hensley2023}. We expect that our results do not qualitatively depend on the form of the size distribution, but the amount of the destroyed interstellar dust is changed. For the power-law distribution with index $-3.5$ \citep{Mathis1977} the short-lived small grains are destroyed rapidly in the SN remnant. However, because the large grains dominate in the total mass of dust, our calculations of the mass of the interstellar dust destroyed in the SN remnant do not change significantly. The upper triangle symbol in Fig.~\ref{fig-md-sat-v} represents the mass of the destroyed interstellar dust with such initial distribution in range $0.003-0.3~\mu$m. For the small-size dominated distributions \citep{Jones1996,Weingartner2001,Hensley2023} a substantial mass fraction of dust would be destroyed. 

\begin{figure}
\includegraphics[width=8.5cm]{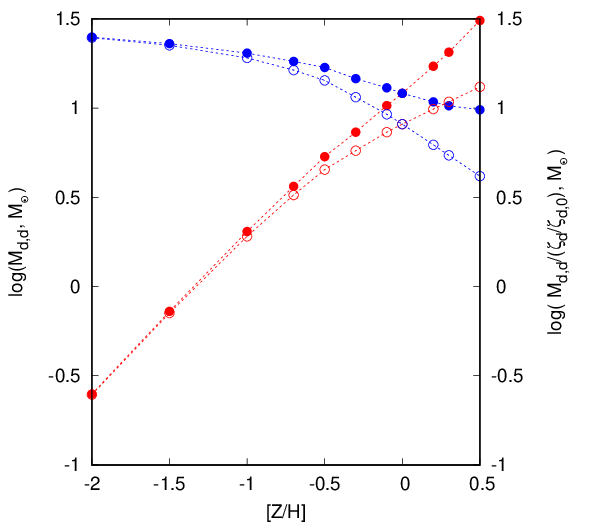}
\caption{
The mass of the interstellar dust destroyed in the bubble expanding in the ambient medium (red line) with density $n=1$~cm$^{-3}$ and metallicity $\rm [Z/H]$. On the right axis this mass is normalized to a factor of the DtG ratio at solar metallicity $\zeta_d/\zeta_{d,0}$ (blue line). Filled symbols correspond to the models without dust cooling, open symbols show the models with dust cooling. These lines present the the runs with the fiducial spatial resolution.
}
\label{fig-md-sat-z}
\end{figure}

\subsection{Dust cooling}

Dust grains are not only destroyed in the hot gas of the SN remnant, but also able to cool it down efficiently  \citep{Ostriker1973,Burke1974,Silk1974,Smith1996,Seok2013,Seok2015,Shchekinov2025jcap}. In the numerical calculations analyzed previously we did not take dust cooling into account. Here we consider its influence. To do this, we account for such cooling in the gasdynamical equations as described in \citet{Dedikov2025irx}. The grains are heated in collisions with electrons \citep{Draine1979a} and emit this energy in the infrared range. We assume thermal equilibrium for the grains \citep[e.g.][]{Dwek1992}. Note that the small grains with size of $a \sim 30$~\AA \ can experience strong temperature fluctuations in the hot gas \citep[e.g.][]{Dwek1986,Drozdov2019}, and their cooling rate is not equal to the rate in the equilibrium. It is fairly suggesting that such small grains are rapidly sputtered, so that their contribution into cooling is negligible. 

Fig.~\ref{fig-md-sat-v} shows the same models as in Fig.~\ref{fig-md-sat}, but taking into account dust cooling. The mass drops in a factor of $\sim 1.7$ for $n\sim 1$~cm$^{-3}$, while for other gas density the decrease is by lower value. The mass remains maximal at $n\sim 1$~cm$^{-3}$ and decreases logarithmically in both more rarefied and denser environments, but with flatter slope. For a power-law grain-size distribution the mass drops about a factor of $\sim 1.4$ for $n\sim 1$~cm$^{-3}$ as well (see the triangle symbols in Fig.~\ref{fig-md-sat-v}). This decrease arises due to the extra cooling of grains in hot gas. In denser environments this decrease becomes smaller. This trend is apparent in decreasing difference between magenta and red lines for higher densities (Fig.~\ref{fig-md-sat-v}).

\subsection{Dependence on metallicity}

Previously, we consider the SN bubble evolution in a gas with solar metallicity. A change of gas metallicity modifies the thermal history of a gas and, consequently, leads to (proportional) alteration of the DtG ratio. Fig.~\ref{fig-md-sat-z} presents the mass of the interstellar dust destroyed in the bubble expanding in the ambient medium with density $n=1$~cm$^{-3}$ for various gas metallicity values $\rm [Z/H]$ (see left axis, filled red symbols). The logarithm of the dust mass depends linearly on gas metallicity: ${\rm log} M_d \sim 0.9{\rm [Z/H]}$, or on the DtG ratio as $M_d \sim \zeta_d$. Such behaviour can be explained by almost linear dependence of the cooling rate $\Lambda(T)$ at $T\sim 10^6$~K on metallicity for $\rm [Z/H]\simgt -2$ \citep[see][]{Gnat2007,v13}. That results in a proportional change in the beginning of radiative phase and the exhausting of grain sputtering. However, the rate $\Lambda$ demonstrates more complex dependence on $(T,Z)$ in hot gas \citep[e.g. Appendix A in][]{v2023}. It follows $\sim T^{0.5}$ at high $T$ for low metallicity and changes its slope to $T^{-0.5}$ with increasing metallicity. This results in longer duration of adiabatic phase in the SN bubble expanding in a gas with lower metallicity. Thus, the dust in the SN remnant evolving in low-metallicity medium is destroyed more efficiently than in case of higher metallicity medium. Here we assume that the DtG ratio is proportional to metallicity. On the right axis of Fig.~\ref{fig-md-sat-z} we show the mass of the interstellar dust normalized to a factor of the DtG ratio at solar metallicity $\zeta_d/\zeta_{d,0}$ (filled blue symbols). It is seen that dust is destroyed with higher efficiency in lower metallicity gas. For $\rm [Z/H]\simlt -1$ there is about two fold increase of the destruction efficiency compared to that of a solar metallicity gas. Fig.~\ref{fig-md-sat-z} presents the dust mass in the models with taking into account dust cooling (open symbols). These extra energy losses facilitate dust surviving in bubble expanding in the ambient medium with $\rm [Z/H]\simgt -0.5$ \citep[see also][]{Shchekinov2025jcap}.

\section{4.~Discussion} \label{sec:discus}

First of all, we should briefly mention other processes related to the dust destruction. Along thermal sputtering there is also kinetic sputtering in the hot gas of the SN bubble \citep{Draine1979b}. It operates in relative gas-grain flows with velocities as high as $\simgt 300$~km/s \citep{Nozawa2006}, which can be found at the inner part of the SN shell. Its contribution rarely exceeds 15\% of the total rate of dust destruction. Another process is shattering in grain-grain collisions \citep{Jones1996,Hirashita2009,Murga2019}. It works in colder ($T\simlt 10^5$~K) gas with velocity dispersion $\sigma_v \simgt 30$~km/s \citep{Murga2019}. A fraction of the interstellar dust contained in such flows is expected to be rather insufficient during the SN remnant evolution \citep{Dedikov2025na}. Moreover, typical timescales for this process are longer than the final time of the numerical calculations considered here. 

Diffuse gas density $n\sim 1$~cm$^{-3}$ at the midplane of the Milky Way disk corresponds to the surface gas density $\Sigma_{gas} \sim 10~\msun$~pc$^{-2}$ \citep[see for details, e.g.][]{Li2017}. According to the Kennicutt-Schmidt relation this is quite moderate value \citep{Kennicutt2012}. However, the destruction of the interstellar dust in SN remnants expanding in an ambient medium with such gas density is the most effective (Fig.~\ref{fig-md-sat-v}). In denser Galactic regions, e.g. the Central Molecular Zone (CMZ) with $\Sigma_{gas} \sim 10^2~\msun$~pc$^{-2}$, the mass of the dust destroyed per SN is expected to be lower. While in the dust-rich ultraluminous galaxies (ULIRGs) the surface density can be more than a factor of $\sim 10^3$ higher. Extrapolating the dependence shown in Fig.~\ref{fig-md-sat-v} for higher density we estimate the dust mass less than $\sim 1~\msun$ for $n\sim 10^3$~cm$^{-3}$. This value is about an order of magnitude lower than that obtained under conditions in a diffuse gas of the Milky Way disk (see Fig.~\ref{fig-md-sat-v}) or estimated from eqn.~\ref{eq:md}. Therefore, compact SN remnants evolving in dense environment are expected to destroy smaller amount of the interstellar dust. Note that SN explosions in dense molecular clouds are commonly considered as compact radio and X-ray sources \citep[e.g.][and others]{Wheeler1980,Shull1980,Chevalier1982,Draine1991,Huang1994,Smith1998,Chevalier1999,Chevalier2001}. Dense gas is typical for an ultraluminous galaxy like Arp 220 \citep{Sakamoto2008,Scoville2015} and massive galaxies at high redshifts \citep{Decarli2018,Decarli2022}, where the starformation is supported at high rate due to powerful galaxy mergers \citep[e.g.][]{Elmegreen2015}. On the other hand, in dust-poor dwarf galaxies both density and metallicity of gas are lower than those in the Milky Way disk. Therefore, the dust destruction rate per SN becomes lower as well. Thus, the dust budget in galaxies \citep[e.g.][]{Mattsson2021,Peroux2023,Shchekinov2025gal} is expected to differ from that based on the simple assumptions \citep[e.g.][]{Draine2009}.

We note that massive stars are usually associated with stellar clusters \citep[e.g.][]{Krumholz2019}. In bubbles driven by multiple SNe explosions the interstellar dust destruction rate per SN is expected to be inhibited compared to that for an isolated SN, because subsequent SNe explosions proceed in almost empty hot bubble, while the shell is dense and cold \citep{vn2026}. A similar scenario occurs when SN explodes in a pre-existing HII region \citep{Martinez2019}.

\section{5.~Conclusion} \label{sec:concl}

In this work we have considered the destruction of the swept-up interstellar dust in the bubble formed by a single SN explosion. Using the 3D hydrodynamic simulations we have calculated how the single-size interstellar dust grains are destroyed in such bubble expanding in a homogeneous medium. Our results can be summarized as follows

\begin{itemize}
\item there are two regimes of the interstellar dust destruction in SN remnants: rapid and almost complete destruction of grains in compact less-massive SN remnants expanding in the dense medium, and gradual and weak destruction in large massive remnants evolving in low-dense environment. In high density gas the sputtering time in the remnant is shorter than the dynamical one, but this period is quite short due to efficient cooling. In rarefied gas the sputtering time exceeds the dynamical one in several times, so the destruction is less efficient, but it continues for longer period;

\item when the time for thermal sputtering is close to the dynamical one, i.e. to the SN remnant age, the mass of the interstellar dust destroyed in the SN remnant reaches maximum value equal to $\sim 13~\msun$ in the ambient gas with density $n\sim 1$~cm$^{-3}$ and solar metallicity, this value is in excellent agreement with the simple estimates by \citet{McKee1989};

\item the mass of the interstellar dust destroyed in the SN remnant decreases both in low and high density environment as the logarithm of the ambient gas density ${\rm log}~n$; using this dependence the mass of destroyed dust is about $5~\msun$ for SN evolving in the ambient medium with density as high as $n \sim 10^2$~cm$^{-3}$, and extrapolating to $n \sim 10^3$~cm$^{-3}$ it drops to $1~\msun$; the latter is about several times smaller than that for SN expanding in the Milky Way diffuse interstellar medium;

\item dust cooling suppresses the interstellar dust destruction, this extra energy losses reduces the SN remnant radius and, as a consequence, the timescale for efficient sputtering. This difference decreases for higher density of the ambient medium;

\item the mass of the dust destroyed in the SN remnant depends linearly on gas metallicity as ${\rm log} M_{d,d} \sim {\rm [Z/H]}$ or, in other words, on the dust-to-gas ratio as $M_{d,d} \sim \zeta_d$, if the ratio is proportional to metallicity; however, the destruction efficiency is higher in low-metallicity environments due to relatively longer adiabatic phase; for ${\rm [Z/H]}\simlt -1$ there is about two fold increase of the destruction efficiency compared to that of a solar metallicity gas; dust cooling can facilitate dust surviving in the ambient medium with $\rm [Z/H]\simgt -0.5$.
\end{itemize}

\section*{ACKNOWLEDGEMENTS}
We thank Yuri A. Shchekinov and Biman B. Nath for many valuable comments, Ilya Khrykin for careful reading of the paper, Svyatoslav Dedikov for discussions, Igor Plavelsky and Denis Yavna for helping with computer hardware that made possible the most time-consuming numerical calculations presented here.

\section*{FUNDING}
This work was supported by ongoing institutional funding. No additional grants to carry out or direct this particular research were obtained.

\section*{CONFLICT OF INTEREST}

The authors of this work declare that they have no conflicts of interest.


\bibliographystyle{aspb1}
\bibliography{p-bib1}

\end{document}